\documentstyle[12pt,aasms4]{article}
\received{}
\accepted{}
\slugcomment{Accepted for publication in The Astronomical Journal}
\lefthead{Eskridge \& Schweitzer}
\righthead{Ursa Minor Dwarf}

\begin{document}

\title{Analysis of a Proper-Motion Selected Sample of Stars in the Ursa Minor 
Dwarf Spheroidal Galaxy}

\author{Paul B.~Eskridge\altaffilmark{1,2} \& Andrea 
E.~Schweitzer\altaffilmark{3}}

\altaffiltext{1}{Department of  Astronomy, The Ohio State University, Columbus,
OH 43210}
\altaffiltext{2}{Department of Physics and Astronomy, Minnesota State 
University, Mankato, MN 56001}
\altaffiltext{3}{Little Thompson Observatory, P.O.~Box 930, Berthoud CO 80513}

\authoremail{eskridge@odin.physastro.mnsu.edu, schweitz@frii.com}

\begin{abstract}
We have studied the stellar population and internal structure of the core of 
the Ursa Minor dwarf spheroidal galaxy, using a sample of stars selected to be
members based on their proper motions.  In agreement with previous studies, we
find Ursa Minor to be dominated by an old, metal-poor stellar population.  A 
small number of stars with high membership probabilities lie redward of the red 
giant branch.  The brightest ($V \la 18$) such stars are known to be Carbon 
stars, rather than metal-rich first-ascent giants.  A number of stars with high 
membership probabilities lie blueward of the red giant branch, and are more 
luminous than the horizontal branch.  We speculate that these are 
post-horizontal branch stars.  There may also be one or two stars in the 
post-AGB phase.  Spectroscopy of the candidate post-HB and post-AGB stars is 
required to determine their nature.  We recover the internal substructure in 
Ursa Minor that has been noted by several authors in the last 15 years.  Using 
a variety of two- and three-dimensional statistical tests, we conclude that 
this substructure is statistically significant at the 0.005 level.  There is no 
evidence that the regions of density excess have stellar populations that 
differ from the main body of Ursa Minor.  The crossing time for a typical 
density excess is only $\sim 5 \times 10^6$ yr.  They are therefore clearly not 
due to intermediate age star-forming bursts.  We conclude that they are instead 
due to tidal interactions between the Galaxy and Ursa Minor.
\end{abstract}


\section{Introduction}

The discovery of the Ursa Minor dwarf spheroidal (dSph) galaxy was announced by 
\markcite{w55}Wilson (1955), although the first comments on its stellar 
populations were by \markcite{baa}Baade (1950).  Ursa Minor is one of four 
dSph satellites of the Galaxy that were identified by direct examination of 
first epoch Palomar Sky Survey plates (\markcite{hnw}Harrington \& Wilson 1950; 
\markcite{w55}Wilson 1955).  The first color-magnitude diagram (CMD) was 
presented by \markcite{vgt}van Agt (1967).  These and other early photographic 
studies led to the conclusion that dSphs were pure population II systems.  
Their stellar populations appeared to be the same as those of Galactic globular 
clusters, but with central stellar densities lower by some six orders of 
magnitude.  Several decades later, Ursa Minor remains the only dSph that 
appears to have a pure pop II stellar population (e.g., \markcite{mnb}Mighell 
\& Burke 1999; \markcite{hgv}Hernandez, Gilmore \& Valls-Gabaud 2000).

\markcite{h64}Hodge (1964) performed the first structural study of Ursa Minor, 
and showed that it is a remarkably elongated system.  He found $e = 0.55 \pm 
0.10$.  This compares very well with the result of \markcite{inh}Irwin \& 
Hatzidimitriou (1995), who found $e = 0.56 \pm 0.05$.  \markcite{h64}Hodge 
(1964) used star counts on Schmidt plates to fit the stellar distribution of 
Ursa Minor with an analytic \markcite{k62}King (1962) model.  His results for
the core and tidal radii are also consistent with the more modern values of
\markcite{inh}Irwin \& Hatzidimitriou (1995).  

The first paper to study the two-dimensional structure of Ursa Minor was 
\markcite{ona}Olszewsky \& Aaronson (1985).  This paper showed evidence for 
non-axisymmetric structure in the stellar distribution of Ursa Minor.  Both the
Fornax (\markcite{h61}Hodge 1961; \markcite{e8b}Eskridge 1988b; 
\markcite{inh}Irwin \& Hatzidimitriou 1995) and Sculptor 
(\markcite{e8a}Eskridge 1988a; \markcite{inh}Irwin \& Hatzidimitriou 1995) 
dSphs are also known to have non-axisymmetric structure in their stellar 
distributions.  Recent work (\markcite{maa}Mart\'{\i}nez-Delgado et al.~2001)
indicates that a substantial population of extra-tidal stars exists surrounding 
Ursa Minor.  It is unclear if the extra-tidal population is in the form of 
tidal tails, or if it is an extended diffuse halo.  The existence of this 
extended structure argues that Ursa Minor is, in fact, not in virial 
equilibrium, and may have great impact on our understanding of the halo dwarfs.

In this paper, we use the proper motion-selected sample of 
\markcite{phd}Schweitzer (1996, see also \markcite{asp}Schweitzer, Cudworth \& 
Majewski 1997), discussed in \S 2, to study the stellar populations and 
structure of the core of Ursa Minor.  In \S 3, we test the hypothesis that Ursa 
Minor is a pure population II system, and seek to either detect any signature 
of a younger stellar population, or to put a numerical limit on such a 
contribution;  In \S 4, we apply a set of statistical tests to the proper 
motion selected sample in order to assess the significance of the sub-structure 
noted in the works above.  In \S 5 we summarize our results, and give 
suggestions for future research.  Following \markcite{mnb}Mighell \& Burke 
(1999), we adopt a distance modulus of $(m-M)_0 = 19.18 \pm 0.12$, and 
foreground reddening and absorption of $E(B-V) = 0.03 \pm 0.01$ and $A_V = 0.09 
\pm 0.03$, giving a linear distance of $69 \pm 4$ kpc.

\section{Data Acquisition and Reduction}

The data for this study are taken from \markcite{phd}Schweitzer (1996).  That 
work provided membership probabilities based on proper motions for stars 
brighter than $V \approx 20.5$ (slightly fainter than the horizontal branch) 
within $\sim 23'$ of the adopted centroid of Ursa Minor: $\alpha_{2000} = 15^h 
08^m\llap.8$, $\delta_{2000} = +67^{\circ} 12'$.  The area surveyed is larger 
than the Ursa Minor core ($r_c = 15'\llap.8 \pm 1'\llap.2$, \markcite{inh}Irwin 
\& Hatzidimitriou 1995).  The catalog has a total of 1533 stars.  Of these, 
1060 have a non-zero membership probability.  The membership-weighted sample is 
867.5 stars.  

\section{Stellar Populations}

Figure 1 shows the $V$ -- $(B-V)$ CMD for the full sample.  The red giant 
branch (RGB) is visible from $V \approx 17$, $(B-V) \approx 1.2$ to $V \approx 
20$, $(B-V) \approx 0.7$.  The horizontal branch (HB) is also visible and, as 
is well known (e.g., \markcite{vgt}van Agt 1967; \markcite{ona}Olszewsky \& 
Aaronson 1985; \markcite{mnb}Mighell \& Burke 1999), dominated by blue HB (BHB) 
stars.  There are also a large number of RR Lyrae variables in the sample.  
However, the CMD is contaminated by a large number of foreground stars.  We can 
use our membership probabilities to produce a much cleaner CMD, shown in Figure 
2.  This figure compares well with Fig.~5 of \markcite{cos}Cudworth, Olszewsky 
\& Schommer (1986), which was based on an earlier proper-motion solution with a 
shorter time baseline.  In Fig.~2, the size of the symbol corresponds to the 
membership probability ($P$), with the smallest dots having $P < 0.1$, and the 
largest circles having $P > 0.9$.  We note that the RGB is now a very 
well-defined feature, and that there is evidence for an asymptotic giant branch 
(AGB) as well.  There are a number of stars with high membership probabilities 
in unusual places of the CMD.  

In region I, there are three stars with $P > 0.5$ that are substantially 
brighter and bluer than the RGB tip.  The joint probability that all three of
these stars are non-members is only $\approx$0.007.  We speculate that one or 
more of them may be evolving across the H-R diagram toward the Planetary Nebula 
stage.  None of them appear to have been studied spectroscopically.  This seems
like a project well worth undertaking.

In region II, there are a dozen stars with $P > 0.5$.  Although one of these 
stars (shown with a cross in Fig.~2a) is known to be a non-member 
(\markcite{hea}Hargreaves et al.~1994), the joint probability that none of them 
are members is $<10^{-13}$.  The members in this region would be post-HB or AGB 
stars.

Finally, in region III, there are 17 stars with $P > 0.5$.  Eight of these are
velocity members based on the spectroscopy of \markcite{hea}Hargreaves et 
al.~(1994), \markcite{aop}Armandroff, Olszweski \& Pryor (1995), and
\markcite{oah}Olszewski, Aaronson \& Hill (1995).  Two are velocity non-members 
according to \markcite{sc2}Shetrone, C\^ot\'e \& Stetson (2001b).  The other 
seven have not been observed spectroscopically.  The joint probability that 
none of these seven stars are members is $P \approx 5 \times 10^{-5}$.  Member 
stars in this region of the CMD are candidate Carbon stars.  Indeed, of the 
eight confirmed members, five are known to be Carbon stars 
(\markcite{aop}Armandroff et al.~1995; \markcite{sc2}Shetrone et al.~2001b).  
One is a Red Giant (\markcite{sc2}Shetrone et al.~2001b), very close to, but 
slightly redward of the RGB tip.  In Fig.~2a, the non-members are shown with
crosses, the Carbon stars with squares, and the red RGB tip star with a 
triangle.

In Table 1, we give the positions, magnitudes, and colors of the stars in all 
three regions of the CMD as an aid for future spectroscopic observations.  We 
note those stars that are known to be velocity members based on the
spectroscopy of \markcite{hea}Hargreaves et al.~(1994), 
\markcite{aop}Armandroff et al.~(1995), \markcite{oah}Olszewski et al.~(1995),
and \markcite{sc2}Shetrone et al.~(2001b).

In Fig.~2b we show the ridge lines for the RGBs of a set of globular clusters,
overlayed on the Ursa Minor CMD.  We use the globular clusters M68
(\markcite{w94}Walker 1994), M55 (\markcite{l77}Lee 1977), NGC 6752 
(\markcite{cns}Cannon \& Stobie 1973), NGC 362 (\markcite{h82}Harris 1982), and
47 Tuc (\markcite{h87}Hesser et al.~1987), with minor corrections as given in
\markcite{dac}Da Costa et al.~(1996), and shifted to account for the distance
and reddening of Ursa Minor.  In Figure 3, we show the probability-weighted CMD
from Fig.~2, along with sets of theoretical isochrones from 
\markcite{gea}Girardi et al.~(2000).  As with the giant branch ridge lines, we
have shifted the isochrones to account for the distance and reddening of Ursa 
Minor.  We show a set of 14 Gyr isochrones for a range of abundances in 
Fig.~3a, and a set of isochrones with $[Fe/H] = -1.7$ for a range of ages in 
Fig.~3b.  As shown by extensive previous work (\markcite{scs}Shetrone, C\^ot\'e 
\& Sargent 2001a, and references therein), Ursa Minor is dominated by an old, 
metal-poor stellar population.  As is clear from Figs.~2b and 3a, there are
proper-motion members of Ursa Minor that fall on the RGB loci of old, 
comparatively metal-rich stellar populations ($[Fe/H] \approx -0.7$).  Recent
spectroscopy of some of the brighter stars in this region of the CMD shows them
to be either non-members or Carbon Stars (\markcite{sc2}Shetrone et al.~2001b). 
At the moment, there is no observational support for the existence of an old, 
but metal-rich stellar population in Ursa Minor. 

Figure 4 shows the $B$- and $V$-band luminosity functions (LFs) of Ursa Minor,
incorporating our membership probabilities.  The error bands are simply root-N
errors of the unweighted number of stars.  That is, any star with a non-zero
membership probability adds one to the sum for error calculations, but adds 
only $P_i$ to the LF.  We applied a Sobel filter (\markcite{mnf}Madore \& 
Freedman 1995; \markcite{lfm}Lee, Freedman \& Madore 1993) in order to search 
for structure in the LFs.  We show an example of the binned differential 
$V$-band LF, and the resulting Sobel-filter output in Figure 5.  The RGB tip is
at $V \approx 16.4$ (or $M_V \approx -2.9$).  There is an apparent gap at $V
\approx 17.5$, and the Red HB stars cause a strong feature at about $V \approx
19$.  No other features are obvious in the Sobel-filter output.  The gap at
$V \approx 17.5$ is apparent in the CMD.  But there is no reason to expect such
a gap at $M_V \approx -1.9$ in an old, metal-poor RGB.  To investigate this
further, we constructed a pure RGB $V$-band LF.  We did this by selecting stars
in the un-numbered region of Fig.~2a, down to $V = 19$.  In Figure 6, we show
the logarithmic integrated $V$-band RGB LF (solid line), along with Poisson 
error bars, computed as above (dotted lines), and a generic RGB LF with a slope
of 0.6 (dashed line).  This shows that the apparent gap at $V \approx 17.5$ is
consistent with being a statistical fluctuation at about the 1-$\sigma$ level.

\section{Tests for Substructure}
\subsection{Background}

Evidence for non-axisymmetric structure in Ursa Minor was first noted by 
\markcite{ona}Olszewsky \& Aaronson (1985).  However, this was based on an 
analysis of the distribution of $\approx$3000 stars in a small ($\approx 3' 
\times 5'$) field.  Subsequent studies by \markcite{inh}Irwin \& Hatzidimitriou 
(1995) and \markcite{kea}Kleyna et al.~(1998) reinforce the notion that there 
is non-axisymmetric clumping in the distribution of stars in Ursa Minor, but 
none of these three studies has been able to make a firm statement on the 
statistical significance of these structures.

\markcite{dea}Demers et al.~(1995) rediscovered one of \markcite{ona}Olszewsky 
\& Aaronson's stellar clumps with their larger-format CCD data, but did not 
realize this at the time, for an accumulation of reasons given in 
\markcite{bnd}Battinelli \& Demers (1999).  \markcite{bnd}Battinelli \& Demers 
(1999) use deep, but very small format, HST/WFPC2 data to argue that the 
density excesses indicated by earlier work are actually due to a ring with a 
central depression.  This is a truly remarkable suggestion, and warrants 
further study by any other means possible.

Our data are not as deep as the \markcite{ona}Olszewsky \& Aaronson (1985) CCD 
data, and cover a smaller field than the \markcite{inh}Irwin \& Hatzidimitriou 
(1995) Schmidt plate scans.  However, our data combine a moderate field of view 
(roughly $40' \times 40'$), with a moderate depth ($V_{lim} \approx 20.5$), and 
with proper-motion based membership probabilities.  Thus the thorny problem of 
foreground subtraction is much less of an issue for our data than any of the 
previous data sets used for structural studies.

In Figure 7a, we show a contour plot of the probability-weighted stellar 
density of our sample.  Figure 7b shows a greyscale map of the density of 
non-members.  We applied a spatial median filter of 9$'$ to the 
probability-weighted membership image.  The filter box-size is roughly the 
minor-axis core radius of Ursa Minor as reported by \markcite{inh}Irwin \& 
Hatzidimitriou (1995).  We experimented with a range of filter kernals, and
found the 9$'$ to be the best at removing the large-scale structure of Ursa 
Minor, without filtering out the regions we are interested in.  After 
generating the filtered image, we then subtracted it from the 
raw image, and show the result in Figure 7c.  Density excesses on either side 
of the centroid are now obvious in the median-filter subtracted data.  We give
the positions of these density excesses in Table 2.  The essential question 
that no previous study has resolved is the following:  Are these substructures 
statistically significant?

\subsection{Analysis}

In many respects, this problem is similar to that of searching for substructure
in clusters of galaxies.  The significant distinction is that one has position 
and radial velocity information for galaxy cluster studies, whilst we have 
position and proper motion information.  Our problem can be made dimensionally 
equivalent to that of galaxy cluster studies by taking membership probability 
as our third measurement.  

\markcite{pea}Pinkney et al.~(1996) compile a set of one- two- and 
three-dimensional tests for substructure of sparse data that they have applied 
to galaxy cluster studies.  We have applied a subset of these tests to our 
sample, using software kindly provided by Jason Pinkney.  Specifically, we use 
the $\beta$, AST, and Lee 2D tests from the two-dimensional substructure tests, 
and the Lee 3D, $\Delta$, $\alpha$, $\alpha_{var}$, and $\epsilon$ tests from 
the three-dimensional substructure tests.  We do not use the 1D tests, as
they search only for substructure in the third variable.  Nor do we use the 
Fourier Elongation Test, as we already know that Ursa Minor is elongated.

In Table 3, we show the results of the 2-D and 3-D tests.  The analysis was
done twice; once using the full data set ($N=1060$), and once using a 
sub-sample of all objects with $P$ within 3$\sigma$ of $\overline P$ ($N=890$).
In practice, the 170 stars removed by the clipping procedure have $P < 0.63$.
All three 2-D tests return highly significant results for substructure using
both the clipped and unclipped samples.  Amongst the 3-D tests, the Lee 3-D
test returns highly significant results for substructure using
both the clipped and unclipped samples.  The $\Delta$, $\alpha$, and 
$\alpha_{var}$ tests consistently give no positive results for substructure,
and the $\epsilon$ test argues for substructure only when applied to the 
clipped data set.  

The $\beta$ test (\markcite{wea}West, Oemler \& Dekel 1988) returns evidence 
for statistically significant subclustering in Ursa Minor on both the clipped 
and unclipped samples.  The $\beta$ statistic is a measure of deviation from 
mirror symmetry.  \markcite{pea}Pinkney et al.~(1996) note that the $\beta$ 
statistic does not measure deviation from circular symmetry, so our result is 
not simply due to the known elongation of Ursa Minor.  Rather, this test argues 
that the distribution of stars about the galaxy center is clumpy, with the 
clumps not being distributed according to mirror symmetry.

The angular seperation test, or AST (\markcite{wea}West et al.~1988) also 
returns evidence for statistically significant subclustering in Ursa Minor on 
both the clipped and unclipped samples.  This test measures the excess of 
small-angle seperations above that expected due to Poisson noise.  
\markcite{wea}West et al.~(1988) examined the behavior of the AST on elongated 
structures, and found that false positives did not occur for ellipticities up
to $e \sim 0.5$ As Ursa Minor has $e=0.56 \pm 0.05$ (\markcite{inh}Irwin \& 
Hatzidimitriou 1995), this is unlikely to be the cause of the positive signal 
we see for the AST.

The Lee 2-D test (\markcite{l2d}Lee 1979) also returns evidence for 
statistically significant subclustering in Ursa Minor on both the clipped and 
unclipped samples.  This test measures the significance of seperating the data 
set into two subclusters.  As with the $\beta$ test, an elongated but smooth 
distribution will not cause the Lee statistic to report a false positive.

The Lee 3-D test (\markcite{fnw}Fitchett \& Webster 1987) is an extension of 
the Lee 2-D test that incorporates variation in the third dimension.  We note 
that if the spatial subclusters have essentially the same distribution in the 
third variable (membership probability, in the case of our Ursa Minor data), 
the Lee 3-D test should give the same result as the Lee 2-D test.  This is the 
result we find for our Ursa Minor data.

Of the other 3-D tests, the $\Delta$, $\alpha$, and $\alpha_{var}$ tests 
consistently give no positive results for substructure.  The $\Delta$ statistic
(\markcite{dns}Dressler \& Shectman 1988) looks for significant variations in 
the third dimension for spatial clumps.  As our third dimension is membership 
probability, the null result we obtain simply says that the spatial subclusters 
(if any) are part of Ursa Minor.  That is, the stars in the spatial subclusters 
have the same distribution of membership probability as do the rest of the Ursa 
Minor stars.  The $\alpha$ (\markcite{wnb}West \& Bothun 1990), and 
$\alpha_{var}$ (\markcite{pea}Pinkney et al.~1996) tests both measure the 
dispersions of the third dimension for spatial clusters, and thus, like the 
$\Delta$ statistic, would not be expected to show significant subclustering in 
the present case.

This brings us to the $\epsilon$ test (\markcite{b93}Bird 1993).  In this
context, $\epsilon$ is a measure of the average number of stars per nearest
neighbor group.  Paradoxically, this test gives strong evidence for 
subclustering on the clipped data set, but no evidence for subsclustering on 
the unclipped data set.  This appears to be due to the diluting effects of the 
low membership probability stars on the test statistic (see discussion in 
\markcite{pea}Pinkney et al.~1996).  When only the high membership probability 
stars are retained, the dispersion in membership probability for any spatial 
subcluster will be much smaller, leading to a more significant result for the 
$\epsilon$ test.

The essential result of this set of tests is that we confirm the previous
evidence for subtructure in Ursa Minor and can now quantify the existence of
this substructure with a statistical confidence of $> 0.995$ for all well-posed 
tests of that structure.  The distribution of member stars in Ursa Minor is not 
well described by a smooth distribution.  In \S 5, we discuss the implications 
of this for the evolution of Ursa Minor.

\section{Summary and Discussion}

We have analysed a proper-motion selected sample of stars in the Ursa Minor
dwarf spheroidal galaxy in order to study the stellar populations, and 
structure of this object.  The data are taken from \markcite{phd}Schweitzer 
(1996), and include stars within $\sim$23$'$ of the Ursa Minor centroid, down 
to a limiting magnitude of $V \approx 20.5$.  The membership-weighted sample is
867.5 stars from a total sample of 1533.

The CMD for our sample compares very well with previous bright-star CMDs for
Ursa Minor (\markcite{vgt}van Agt 1967; Cudworth et al.~1986), showing the
dominant old, metal-poor stellar population.  Comparison with globular cluster
RGB ridge lines and model isochrones indicates a metallicity of $[Fe/H] \approx 
-1.8$ to $-2$, consistent with the spectroscopic result of $[Fe/H] = -1.90 \pm 
0.11$ from \markcite{scs}Shetrone et al.~(2001a).  There are a number of 
high-probability members in unusual parts of the CMD.  We provide position and 
magnitude data for these stars as an aid for future spectroscopic observations. 
The $B$- and $V$-band LFs of our sample are consistent with a single, dominant, 
old, metal-poor stellar population.

We have applied a number of tests for substructure to our sample in order to 
evaluate the statistical significance of the well-known clumps of stars in Ursa
Minor (\markcite{ona}Olszewsky \& Aaronson 1985; \markcite{inh}Irwin \& 
Hatzidimitriou 1995; \markcite{dea}Demers et al.~1995; \markcite{kea}Kleyna et 
al.~1998; \markcite{bnd}Battinelli \& Demers 1999).  These tests are drawn from
the study of \markcite{pea}Pinkney et al.~(1996), who compiled a set of one- 
two- and three-dimensional tests for substructure in sparse data for galaxy 
cluster studies.  When the detailed assumptions and behaviour of these tests 
are taken into account, we conclude that the hyphothesis that Ursa Minor has 
statistically significant substructure is strongly supported (see Table 3).  We 
also employed a test devised by \markcite{m2k}Mighell (2001) to search for
structure in sparse 2-D Poisson-distributed data.  The test is a modification 
of the $\chi^2$ statistic.  Given a model, the test reports the significance of 
the deviation of the data from the model.  As the test is well-defined for
integers, we have used the sample of all stars with $P \ge 0.9$ of being
members of Ursa Minor.  Our model is a simple elliptical profile fit to these
data, with ellipticity, and position angle as free parameters.  The test
rejects the null hypothesis that the data follow the model at the 95\%
confidence level.  Thus, this test is consistent with the results from the
clustering statistics discussed in \S 4, and indicates that the distribution of 
stars in Ursa Minor is not well-fit with a smooth model.  

We do not see evidence of the structure reported by \markcite{bnd}Battinelli \& 
Demers (1999).  The lack of agreement between our study and theirs is likely
due to the very different strengths and weaknesses of the data for the two
studies.  The \markcite{bnd}Battinelli \& Demers (1999) study is based on very
deep data covering a very small field of view (a single WFPC2 frame), whereas
our data are very shallow, but cover a much larger field of view.  A proper 
test of the results of \markcite{bnd}Battinelli \& Demers (1999) would require
much deeper wide-field data than we have at our disposal.

The angular size of the lumps is roughly 3$'$ (see Figure 7c).  This is about
20\% of the size of the core radius.  The measured velocity dispersion of Ursa 
Minor is roughly 10 km/sec ($10.4 \pm 0.9$ km/sec according to 
\markcite{aop}Armandroff et al.~1995).  For our adopted distance of $69 \pm 4$ 
kpc, this implies a lump crossing time of only $\sim 5 \times 10^6$ yr.  
However, neither the CMD, nor the $V$-band LF of stars in the lumps shows any 
evidence for differences in the stellar populations of the lumps and the bulk 
of Ursa Minor.  The lumps are clearly not due to young (or even 
intermediate-age) star formation events.  We speculate that they are instead 
due to tidal stretching from the gravitational interaction between the Galaxy 
and Ursa Minor.  Features like the Ursa Minor lumps appear in some numerical 
simulations of such interactions (e.g., \markcite{nak}Kroupa 1997; 
\markcite{knk}Klessen \& Kroupa 1998).  However, the evidence for such 
dynamically induced structure and for extra-tidal stars associated with Ursa 
Minor (\markcite{maa}Mart\'{\i}nez-Delgado et al.~2001) is unlikely to provide 
an escape for the large mass to light ratio for Ursa Minor that is implied by 
its velocity dispersion, as numerical studies indicate that the observed 
velocity dispersions of disrupting dwarfs do not significantly exceed their 
viral values until just before their complete dispersal (e.g., 
\markcite{oea}Oh, Lin \& Aarseth 1995; \markcite{pna}Piatek \& Pryor 1995).  

It would be very interesting to obtain spectra of the remaining 
high-probability member stars in the outlying regions of the CMD.  The stars in 
region I are bright enough ($V < 16.2$) that spectroscopy with a 4m class 
telescope should be able to determine their membership, and their nature.  All 
the high-probability stars in regions II and III that have not yet been 
observed spectroscopically are fainter than $V \approx 18$, and thus become 
difficult targets for 4m-class spectroscopy.  Determining the membership and 
nature of these stars would be an excellent project for the HET, the upgraded
MMT, or the Keck telescope, or, in the future, for the Gemini telescope.  Very 
deep photometry of Ursa Minor, both in the lumps and in the general field, 
would provide a much better probe of the stellar populations than the data 
currently available.  The main problem with the WFPC2 data is that the field of 
view is so small.  Observations of a number of fields in the halo dwarf 
spheroidals with the MOSAIC Imager on the Mayall and Blanco 4m telescopes, and 
with the forethcoming Advanced Camera for Surveys will put our understanding of 
the details of the star formation histories of the halo dwarfs on a much better 
observational footing than is currently possible.  Finally, there is now clear 
evidence that roughly half of the halo dSphs have internal substructure.  The 
lesson of the Sagittarius dSphs is that interactions with the Galaxy can cause 
distortions in the distributions of stars in the halo dwarfs.  The current 
observational situation presents a challenge to theorists that has the 
potential to lead to dramatic improvements in our understanding of the role of 
tidal stripping in the dissolution of dwarf systems, and the building of the 
Galactic halo.

\acknowledgments

We thank Jason Pinkney for use of his clustering statistics software, and
extensive consulation on its care and feeding.  Ken Mighell provided us with 
the results for his 2-D Poisson statistic, and with several very useful 
discussions and comments.  PBE would like to thank the Department of Physics 
and Astronomy at The University of Alabama, and the Department of Astronomy at 
The Ohio State University, and their members for their intellectual and 
financial support during the course of this project.  PBE would also like to 
thank Barry Madore for pointing out what should have been the obvious.  AES
thanks Kyle Cudworth for the use of his proper motion reduction software during
her dissertation work, as well as the Department of Astronomy at The University
of Wisconsin.  We have made extensive use of both NED, the NASA/IPAC 
Extragalactic Database (which is operated by the Jet Propulsion Laboratory, 
California Institute of Technology, under contract with the National 
Aeronautics and Space Administration), and the ADS (NASA's Astrophysics Data 
System Abstract Service).

\newpage

\baselineskip12pt
\tolerance=500

\def\tabrule{\noalign{\hrule}}
\def\pz{\phantom{0}}
\def\pb{\phantom{-}}
\def\pd{\phantom{.}}
\def\po{\phantom{1}}
\

\centerline{Table 1 - CMD Outliers with $P > 0.5$}
\vskip0.2cm

\newbox\tablebox
\setbox\tablebox = \vbox {

\halign{\hfil#\pz&\hfil#\pz\hfil&\hfil#\pz\hfil&\hfil#\pz\hfil&\hfil#\pz&\hfil#
\pz&\pz\hfil#\hfil\pz&\pz\hfil#\hfil\pz&\hfil#\pz&\pz#\pz&\pz#\pz\hfil\cr
\tabrule
\noalign{\vskip0.1cm}
\tabrule
\noalign{\vskip0.1cm}

ID & V & (B-V) & $P$ & X$\pz$ & Y$\pz$ & RA & DEC & COS$^1$ & Irwin$^2$ & 
vel.~mem$^3$ \cr
 & & & & $''\pz$ & $''\pz$ & (2000) & (2000) & & & \cr
\noalign{\vskip0.1cm}
\tabrule
\noalign{\vskip0.2cm}
746 & 16.17 & 0.74 & 0.51 & $-$3.3 & 745.2 & 15 08 51.60 & 67 25 11.5 & & & \cr
783 & 15.86 & 0.71 & 0.96 & $-$564.4 & 465.5 & 15 07 14.50 & 67 20 29.9 & & & 
\cr
1406 & 16.09 & 1.05 & 0.67 & $-$513.2 & $-$99.4 & 15 07 23.94 & 67 11 05.4 & & 
& \cr
\noalign{\vskip0.1cm}
\tabrule
\noalign{\vskip0.2cm}
177 & 18.48 & 0.43 & 0.96 & 797.5 & 915.4 & 15 11 10.90 & 67 27 58.0 & & & \cr
253 & 18.76 & 0.15 & 0.80 & $-$260.3 & 1300.1 & 15 08 06.68 & 67 34 26.0 & & & 
\cr
296 & 18.81 & 0.22 & 0.96 & $-$977.9 & 947.1 & 15 06 02.00 & 67 28 27.8 & & & 
\cr
395 & 19.23 & 0.15 & 0.91 & 23.9 & 1202.8 & 15 08 56.22 & 66 52 43.5 & & & \cr
690 & 18.16 & 0.21 & 0.91 & 378.3 & 907.0 & 15 09 57.97 & 67 27 52.4 & & & \cr
916 & 19.34 & 0.19 & 0.97 & $-$676.6 & $-$606.4 & 15 06 56.53 & 67 02 37.3 & & 
& \cr
949 & 19.00 & 0.14 & 0.88 & $-$644.3 & $-$801.3 & 15 07 02.28 & 66 59 22.7 & & 
& \cr
1089 & 17.92 & 0.40 & 0.91 & 673.5 & 21.6 & 15 10 48.12 & 67 13 05.2 & & & \cr
1259 & 19.01 & 0.16 & 0.93 & 444.3 & 593.8 & 15 10 09.17 & 67 22 38.9 & & & \cr
1537 & 17.53 & 0.88 & 0.84 & $-$113.9 & $-$496.0 & 15 08 32.68 & 67 04 30.2 & 
442 & 37482 & no \cr
1600 & 19.10 & 0.24 & 0.98 & 63.8 & $-$157.9 & 15 09 03.14 & 67 10 08.4 & & & 
\cr
1710 & 18.45 & 0.54 & 0.83 & 136.6 & 341.1 & 15 09 15.78 & 67 18 27.2 & & & \cr
\noalign{\vskip0.1cm}
\tabrule
\noalign{\vskip0.2cm}
152 & 17.62 & 1.33 & 0.74 & 1055.7 & 748.7 & 15 11 55.47 & 67 25 08.5 & & 30614 
& yes,C \cr
251 & 18.07 & 1.42 & 0.91 & $-$445.2 & 1315.8 & 15 07 34.37 & 67 34 40.9 & & & 
no \cr
368 & 19.29 & 1.06 & 0.52 & $-$431.1 & $-$998.1 & 15 07 38.81 & 66 56 07.1 & & 
& \cr
440 & 19.13 & 1.32 & 0.54 & 187.6 & $-$809.1 & 15 09 24.17 & 66 59 16.9 & & & 
\cr
505 & 18.95 & 1.17 & 0.56 & 1052.9 & $-$8.4 & 15 11 53.38 & 67 12 31.5 & & & \cr
536 & 17.58 & 1.39 & 0.88 & 950.2 & 326.1 & 15 11 36.35 & 67 18 07.1 & & 32961 
& yes,C \cr
829 & 17.94 & 1.42 & 0.75 & $-$759.9 & $-$94.6 & 15 06 41.52 & 67 11 08.3 & & & 
no \cr
979 & 18.45 & 1.26 & 0.96 & $-$160.8 & $-$919.4 & 15 08 24.79 & 66 57 26.8 & & 
& \cr
1008 & 19.65 & 1.01 & 0.81 & 221.2 & $-$552.0 & 15 09 30.01 & 67 03 34.0 & & & 
\cr
1128 & 17.15 & 1.38 & 0.66 & 384.9 & 214.1 & 15 09 58.59 & 67 16 19.5 & 82 & 
33521 & yes \cr
1167 & 18.05 & 1.34 & 0.87 & 574.7 & 381.7 & 15 10 31.54 & 67 19 06.1 & 122 & 
32613 & yes,C \cr
1363 & 18.73 & 1.11 & 0.78 & $-$303.9 & 162.8 & 15 07 59.77 & 67 15 28.6 & & & 
\cr
1545 & 17.26 & 1.38 & 0.66 & $-$96.5 & $-$545.2 & 15 08 35.67 & 67 03 41.1 & & 
37759 & yes,C \cr
1677 & 16.54 & 1.56 & 0.98 & 125.1 & 166.8 & 15 09 13.74 & 67 15 33.0 & 60 & 
33767 & yes \cr
1721 & 18.48 & 1.44 & 0.67 & $-$65.3 & 286.7 & 15 08 40.90 & 67 17 33.0 & & & 
\cr
1846 & 16.91 & 1.57 & 0.99 & $-$145.6 & $-$158.9 & 15 08 27.16 & 67 10 07.3 & 
347 & 35606 & yes,R \cr
1859 & 17.92 & 1.18 & 0.94 & $-$251.5 & $-$204.6 & 15 08 08.98 & 67 09 21.3 & & 
35869 & yes,C \cr
\noalign{\vskip0.2cm}
\tabrule
}
Notes: \hfill\break
1:  Numbers from \markcite{cos}Cudworth et al.~(1986). \hfill\break
2:  Numbers from M.~Irwin (unpublished). \hfill\break
3:  Velocity membership status from \markcite{hea}Hargreaves et al.~(1994), 
\markcite{aop}Armandroff et al.~(1995), \markcite{oah}Olszewski et al.~(1995),
or \markcite{sc2}Shetrone et al.(2001).  C: Carbon stars according to 
\markcite{aop}Armandroff et al.~(1995), or \markcite{sc2}Shetrone et al.(2001). 
R:  RGB star according to \markcite{sc2}Shetrone et al.(2001).
}
\centerline{ \box\tablebox}

\newpage

{
\baselineskip12pt
\tolerance=500

\def\tabrule{\noalign{\hrule}}
\def\pz{\phantom{0}}
\def\pb{\phantom{-}}
\def\pd{\phantom{.}}
\def\po{\phantom{1}}
\

\centerline{Table 2 - Lump Positions}
\vskip0.2cm

\newbox\tablebox
\setbox\tablebox = \vbox {

\halign{\hfil#\hfil&\hfil\pz\pz#\pz\pz\hfil&\hfil\pz\pz#\pz\pz\hfil\cr
\tabrule
\noalign{\vskip0.1cm}
\tabrule
\noalign{\vskip0.1cm}

Number & RA (2000) & Dec (2000) \cr
& $hhmmss$ & $~^\circ~^\circ ~^{'}~^{'} ~^{''} ~^{''}$ \cr
\noalign{\vskip0.1cm}
\tabrule
\noalign{\vskip0.2cm}
1 & 15 08 00 & 67 06 30 \cr
2 & 15 08 30 & 67 10 20 \cr
3 & 15 08 40 & 67 12 30 \cr
4 & 15 09 10 & 67 11 40 \cr
5 & 15 10 10 & 67 21 50 \cr
6 & 15 10 30 & 67 19 50 \cr
\noalign{\vskip0.2cm}
\tabrule
}
}
\centerline{ \box\tablebox}
}

{
\baselineskip12pt
\tolerance=500

\def\tabrule{\noalign{\hrule}}
\def\pz{\phantom{0}}
\def\pb{\phantom{-}}
\def\pd{\phantom{.}}
\def\po{\phantom{1}}
\

\centerline{Table 3 - Substructure Test Results}
\vskip0.2cm

\newbox\tablebox
\setbox\tablebox = \vbox {

\halign{\pz\pz#\pz\pz&\hfil\pz\pz#\pz\pz\hfil&\hfil\pz\pz#\pz\pz\hfil&\hfil\pz
\pz#\pz\pz\hfil\cr
\tabrule
\noalign{\vskip0.1cm}
\tabrule
\noalign{\vskip0.1cm}

 \ & Test & Statistic & Probability \cr
\noalign{\vskip0.1cm}
\tabrule
\noalign{\vskip0.2cm}
Full Sample & $\beta$ & 11.3 & 0.000 \cr
 (N=1060) & AST & 13.5 & 0.000 \cr
 \ & Lee 2D & 1.758 & 0.000 \cr
 \ & Lee 3D & 1.339 & 0.000 \cr
 \ & $\Delta$ & 10637 & 1.000 \cr
 \ & $\alpha$ & 181.188 & 0.898 \cr
 \ & $\alpha_{var}$ & 182.320 & 0.846 \cr
 \ & $\epsilon$ & 1.57E11 & 0.427 \cr
\noalign{\vskip0.1cm}
\tabrule
\noalign{\vskip0.2cm}
Subsample & $\beta$ & 8.2 & 0.006 \cr
 (N=890) & AST & 11.9 & 0.000 \cr
 \ & Lee 2D & 1.820 & 0.000 \cr
 \ & Lee 3D & 1.369 & 0.000 \cr
 \ & $\Delta$ & 3726 & 0.409 \cr
 \ & $\alpha$ & 203.355 & 0.756 \cr
 \ & $\alpha_{var}$ & 203.086 & 0.756 \cr
 \ & $\epsilon$ & 6.96E10 & 0.002 \cr
\noalign{\vskip0.2cm}
\tabrule
}
}
\centerline{ \box\tablebox}
}

\newpage

\figcaption{$V$ -- $(B-V)$ CMD of the full Ursa Minor sample.}

\figcaption{$V$ -- $(B-V)$ CMD incorporating probability weights.  The smallest
solid dots have membership probabilities of $P < 0.1$.  The largest open 
circles have membership probabilities of $P > 0.9$.  Stars with $P > 0.5$ are
shown as open symbols.  The solid lines demark regions of the CMD, labelled 
with roman numerals I--III, from which the stars listed in Table 1 were drawn.
a). The squares show spectroscopically confirmed member Carbon stars, the 
triangle shows the spectroscopically confirmed member RGB star, the crosses 
show the spectroscopically confirmed non-members.  b). The curves show globular 
cluster RGB ridge lines taken from the literature.  The clusters shown are 47 
Tuc (dotted line), NGC 362 (short dashed line), NGC 6752 (long dashed line), 
M55 (dot - short dash line), and M 68 (dot - long dash line).}

\figcaption{$V$ -- $(B-V)$ CMD from figure 2, with isochrones from Girardi et 
al.~(2000) overlayed.  a) 14 Gyr isochrones with $[Fe/H] = -1.7$ (solid line),
-1.3 (dotted line), -0.7 (short-dashed line), -0.4 (long-dashed line), 0 
(short-dashed-dotted line) and +0.2 (long-dashed-dotted line).  b) $[Fe/H] = 
-1.7$ isochrones with ages of 18 Gyr (solid line), 14 Gyr (dotted line), 11 Gyr
(short-dashed line), 9 Gyr (long-dashed line), 7 Gyr (short-dashed-dotted line)
and 5.6 Gyr ((long-dashed-dotted line).}

\figcaption{Logarithmic probability-weighted integrated LF of Ursa Minor.  The
solid lines are the LFs, and the dashed lines represent root-N error bounds of
the unweighted number of stars.  a) $B$-band LF.  b) $V$-band LF.}

\figcaption{a) Differential $V$-band LF with 0.2 magnitude binning. b) 
Sobel-filter output of the data shown in a).}

\figcaption{Logarithmic $V$-band integrated LF of the RGB of Ursa Minor.  The
solid line is the LF, the dashed lines represent root-N error bounds of the 
unweighted number of stars, the dashed line is a model LF with a power law 
index of 0.6.}

\figcaption{Density maps of stars in the Ursa Minor sample.  The origin is the
adopted center of Ursa Minor.  North is up, East to the left.  a) Contour plot
of the probability-weighted density of Ursa Minor stars binned into a 1$' 
\times 1'$ image.  b) A grey-scale plot of the non-members at the same 
resolution.  c) A contour plot of the residuals of the membership-weighted 
array minus a median-filtered version of the array, with a 9$'$ filter box.}

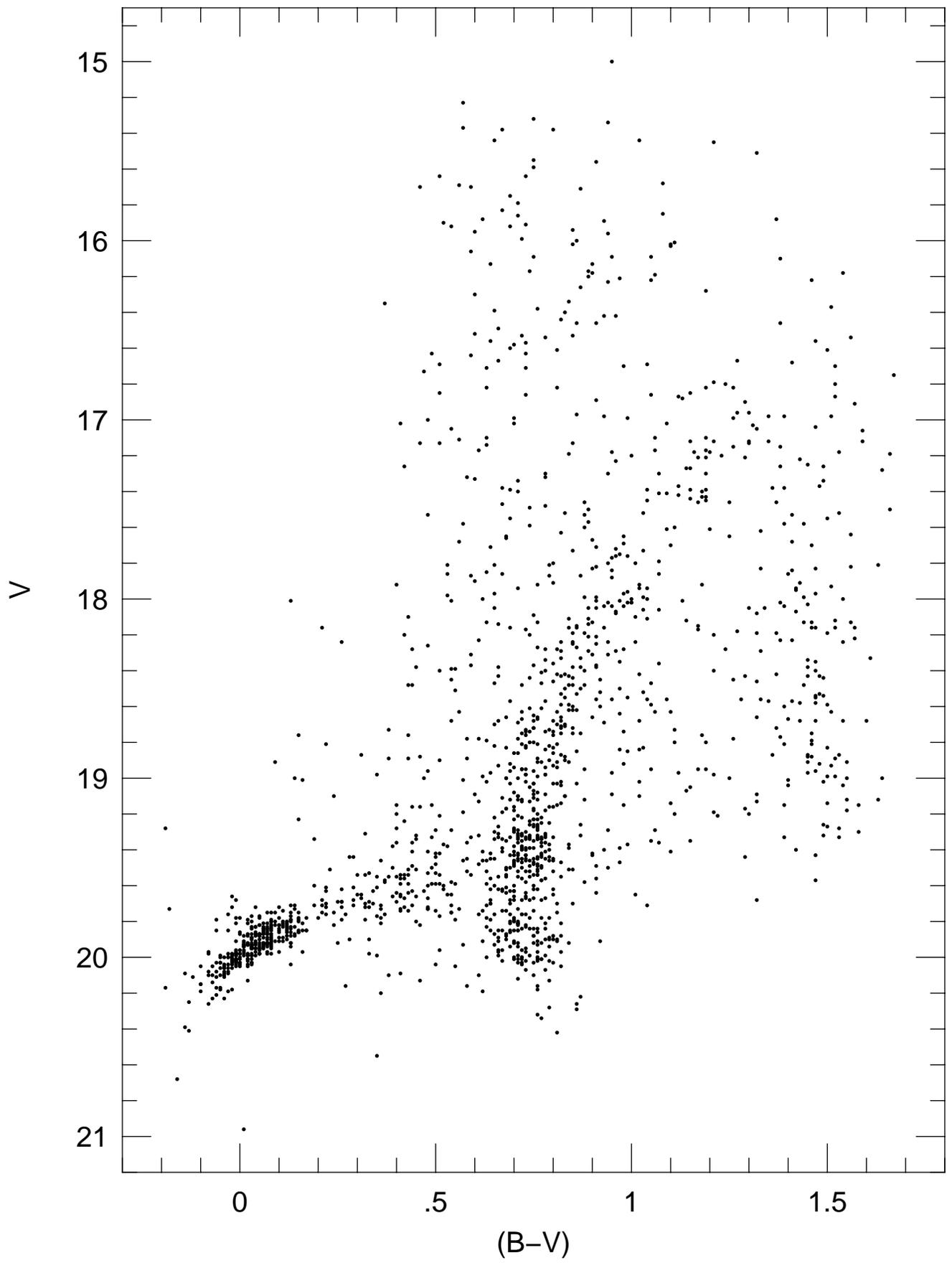
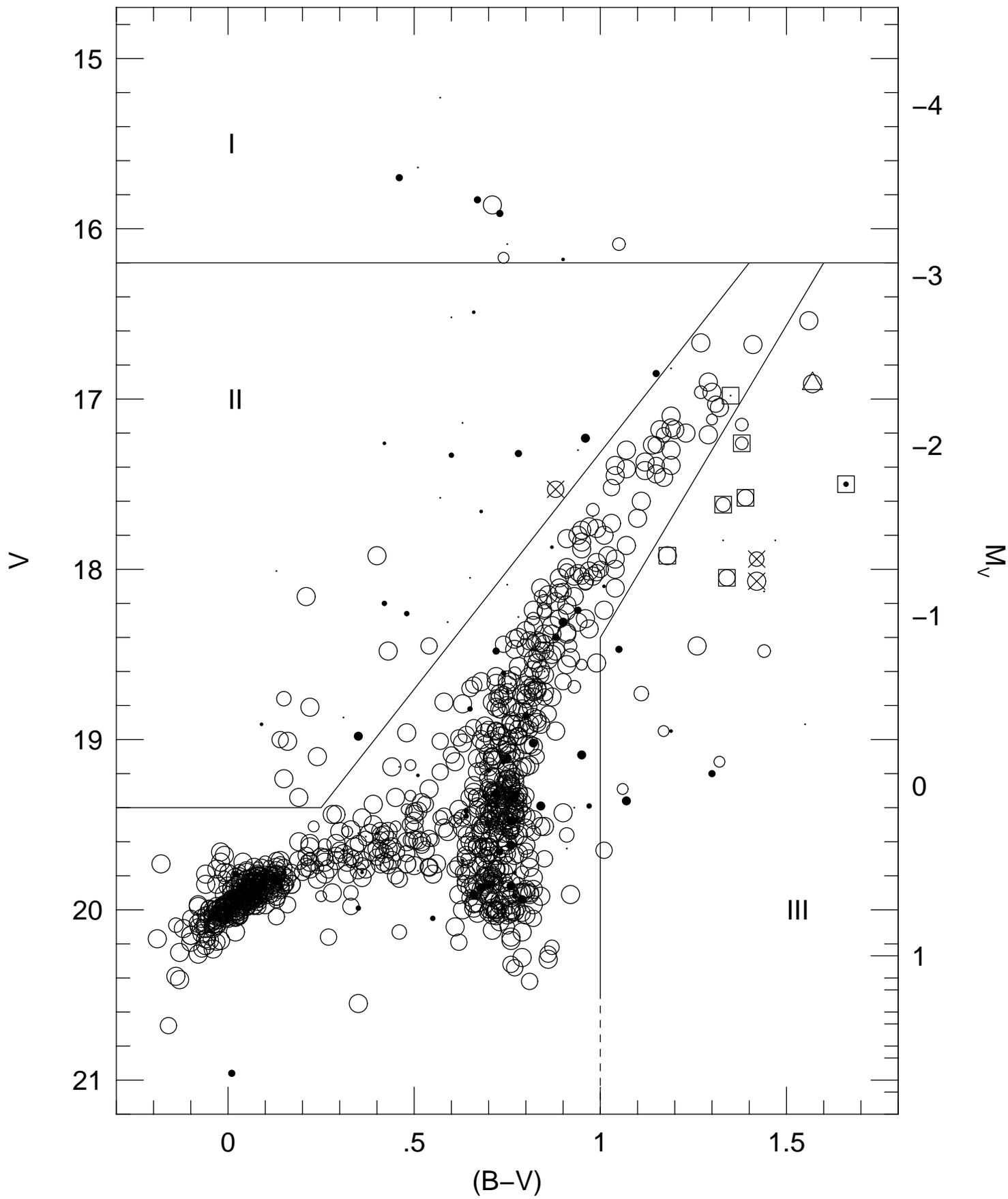
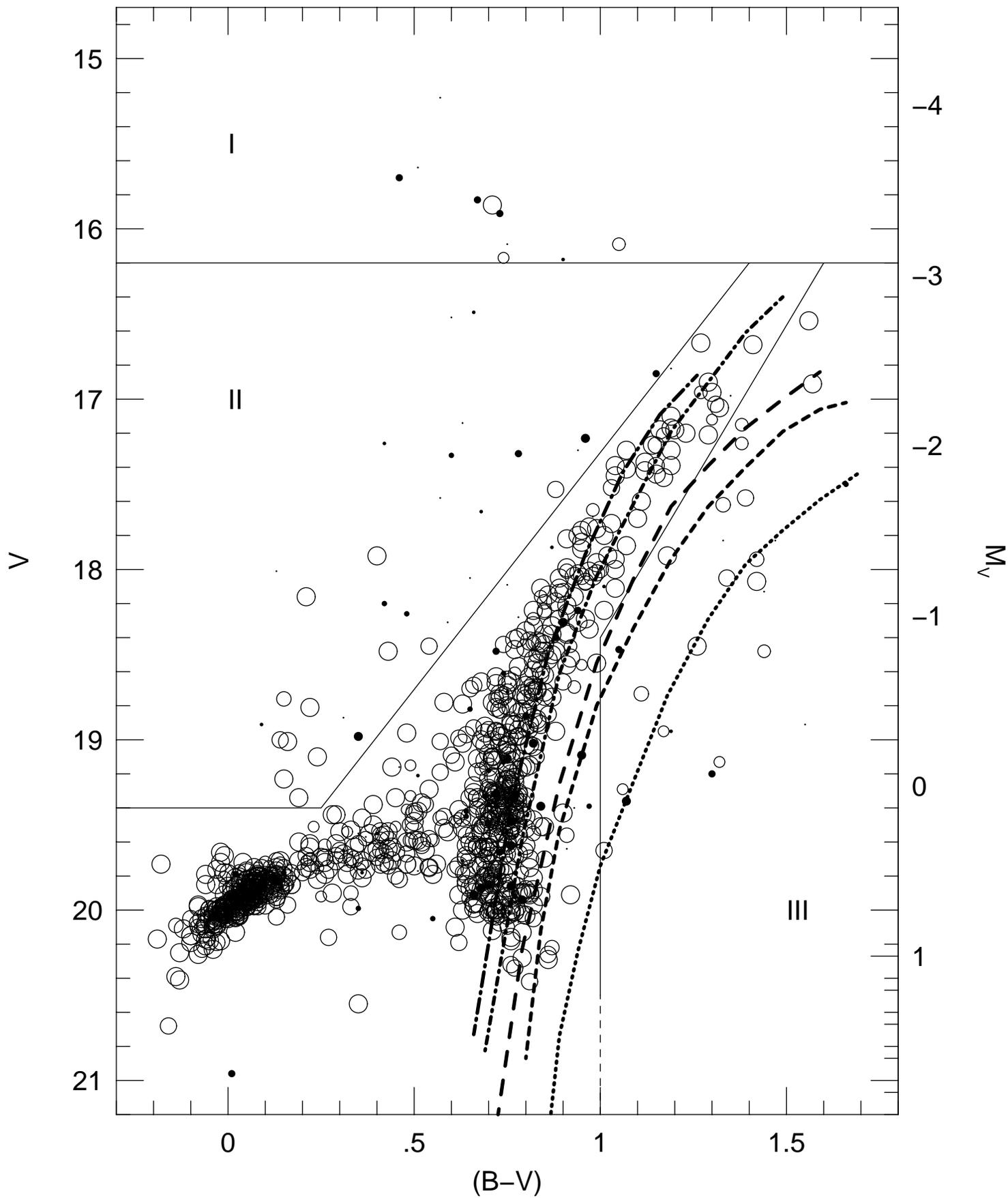
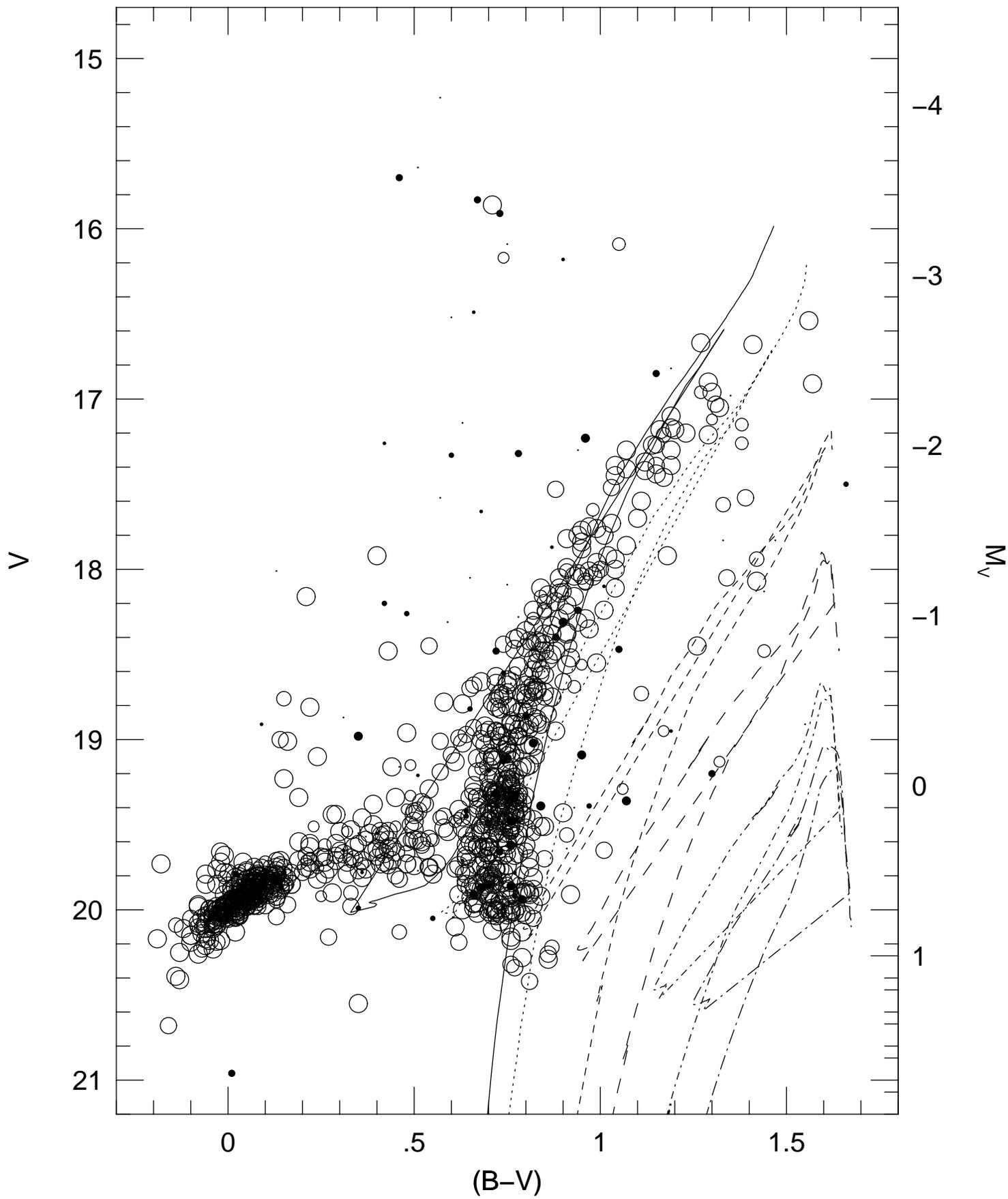
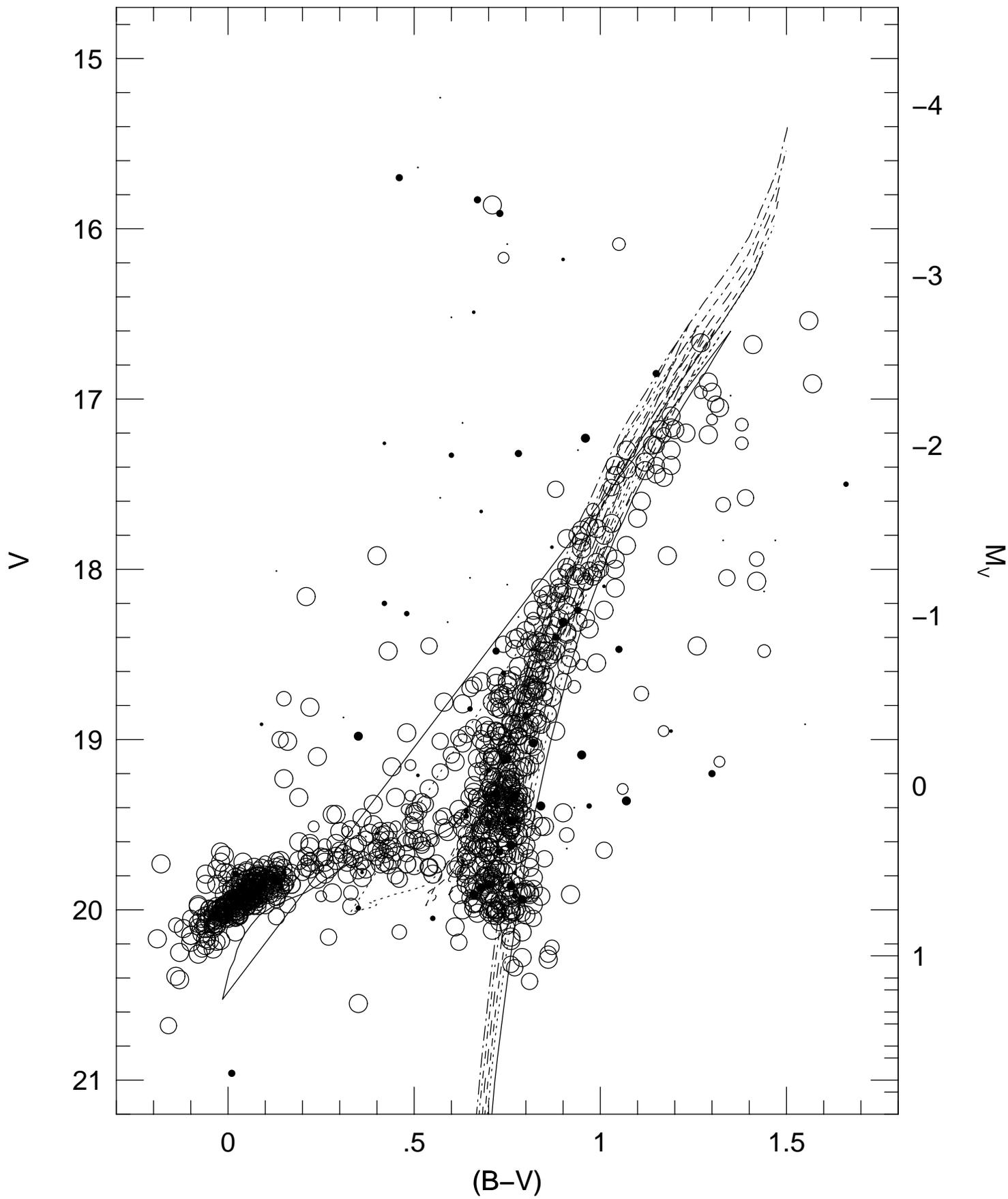
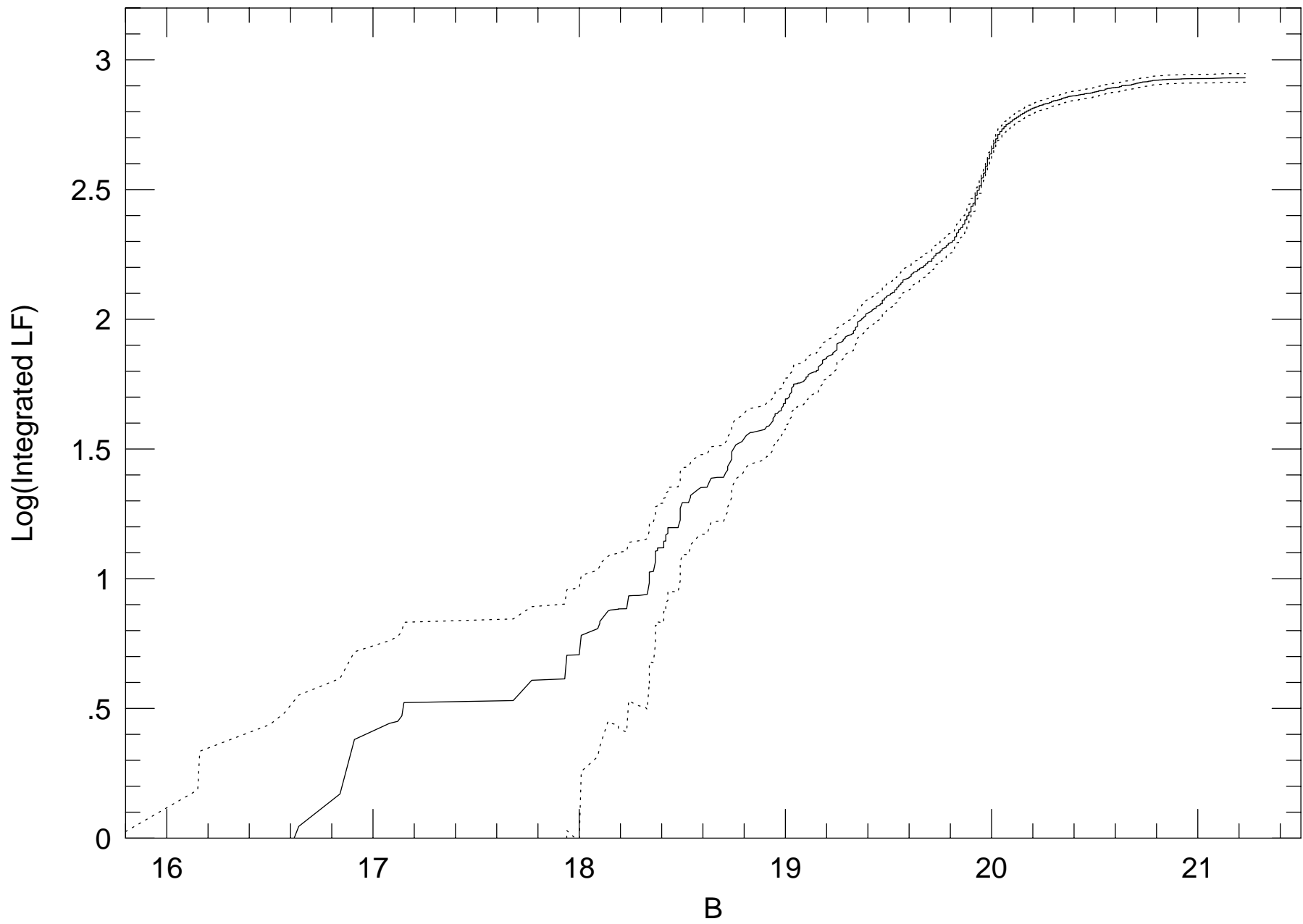
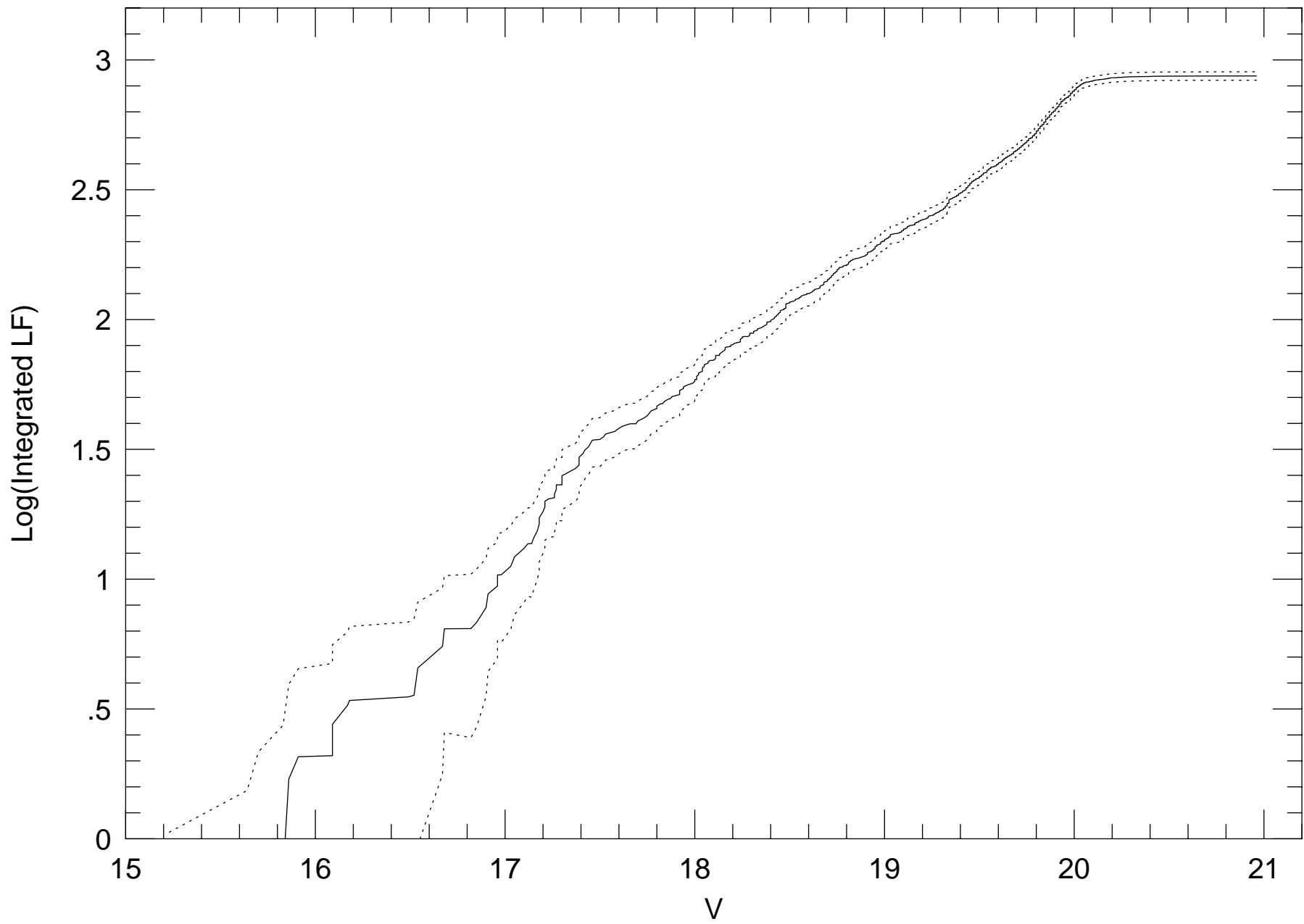
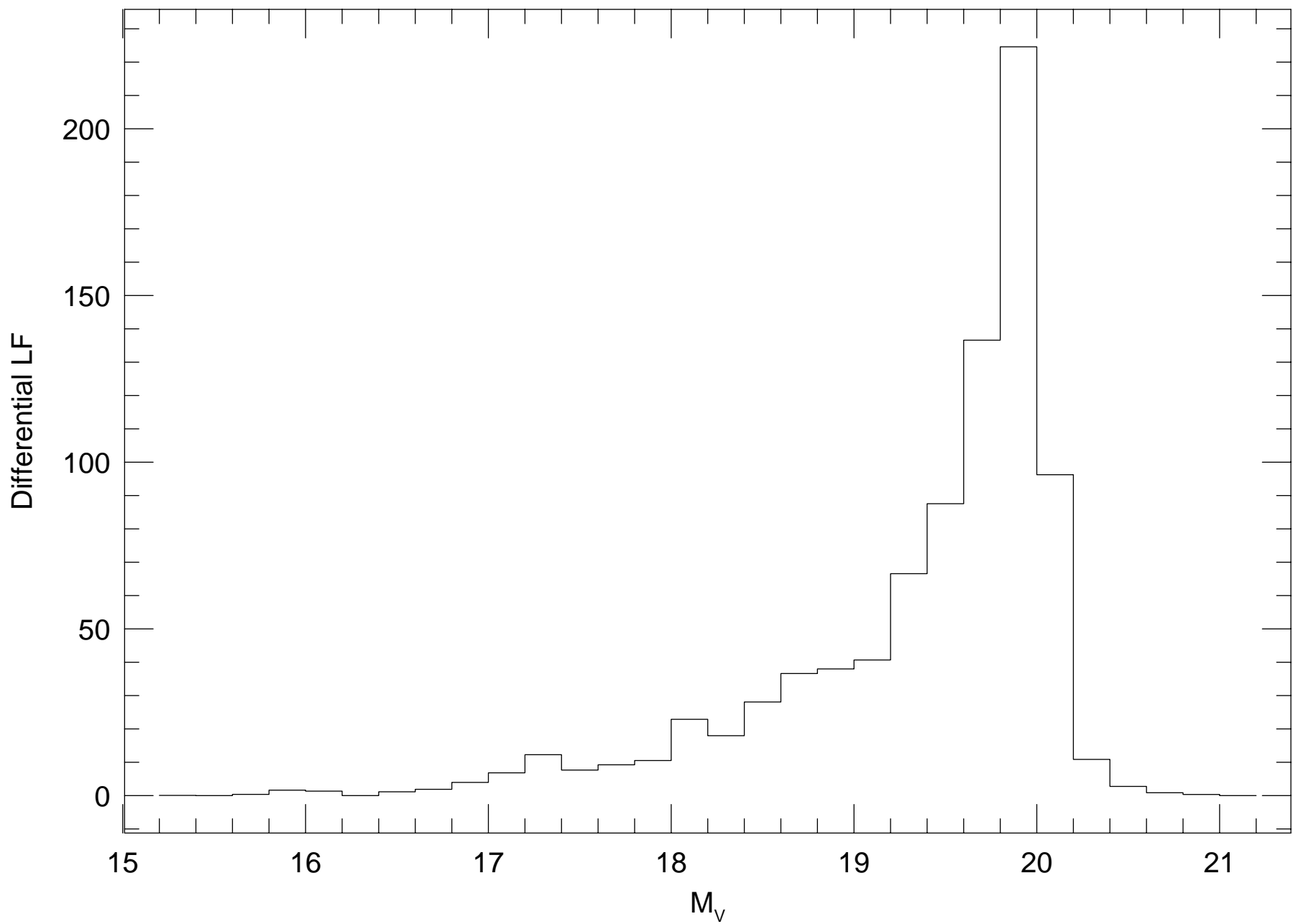
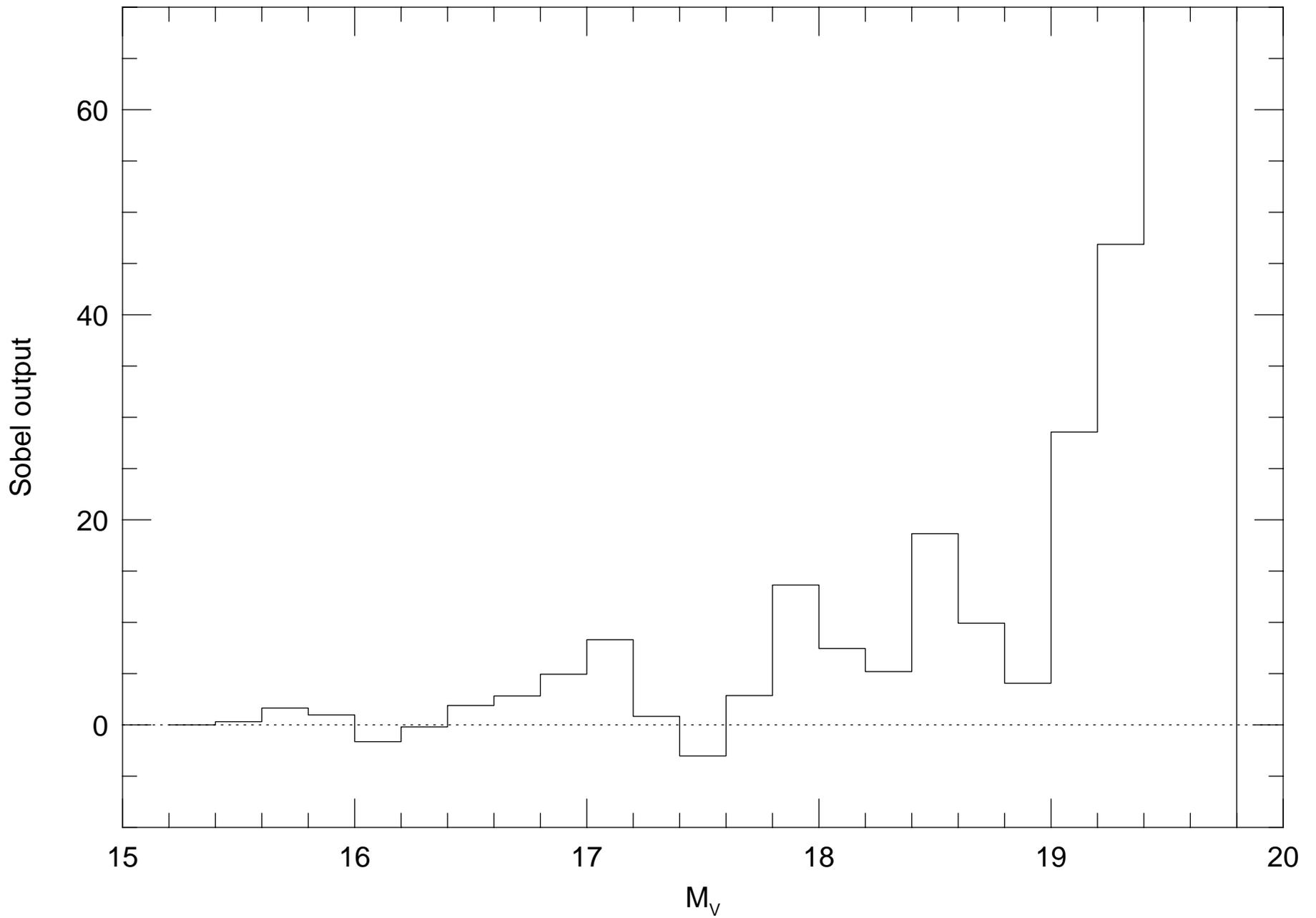
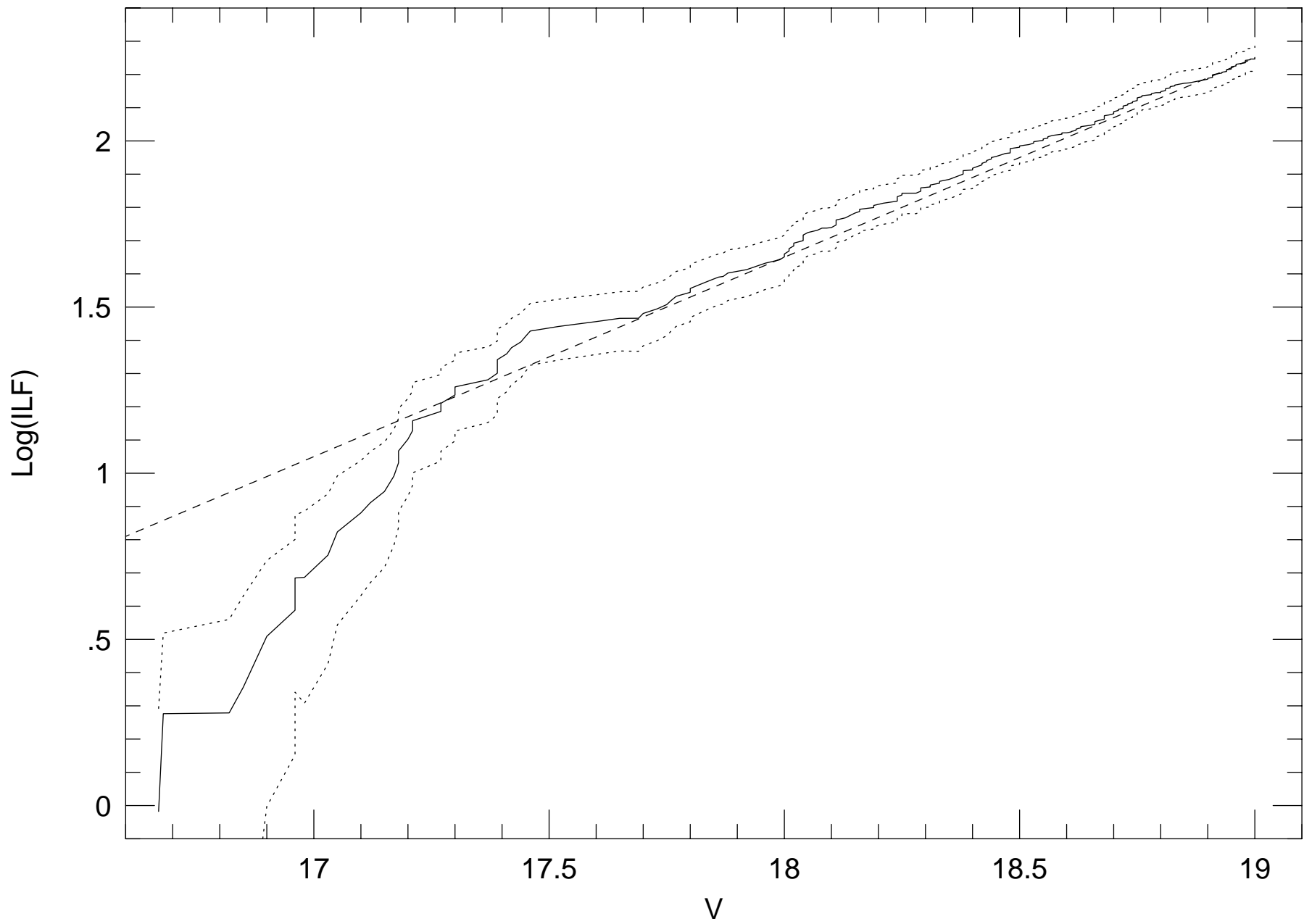
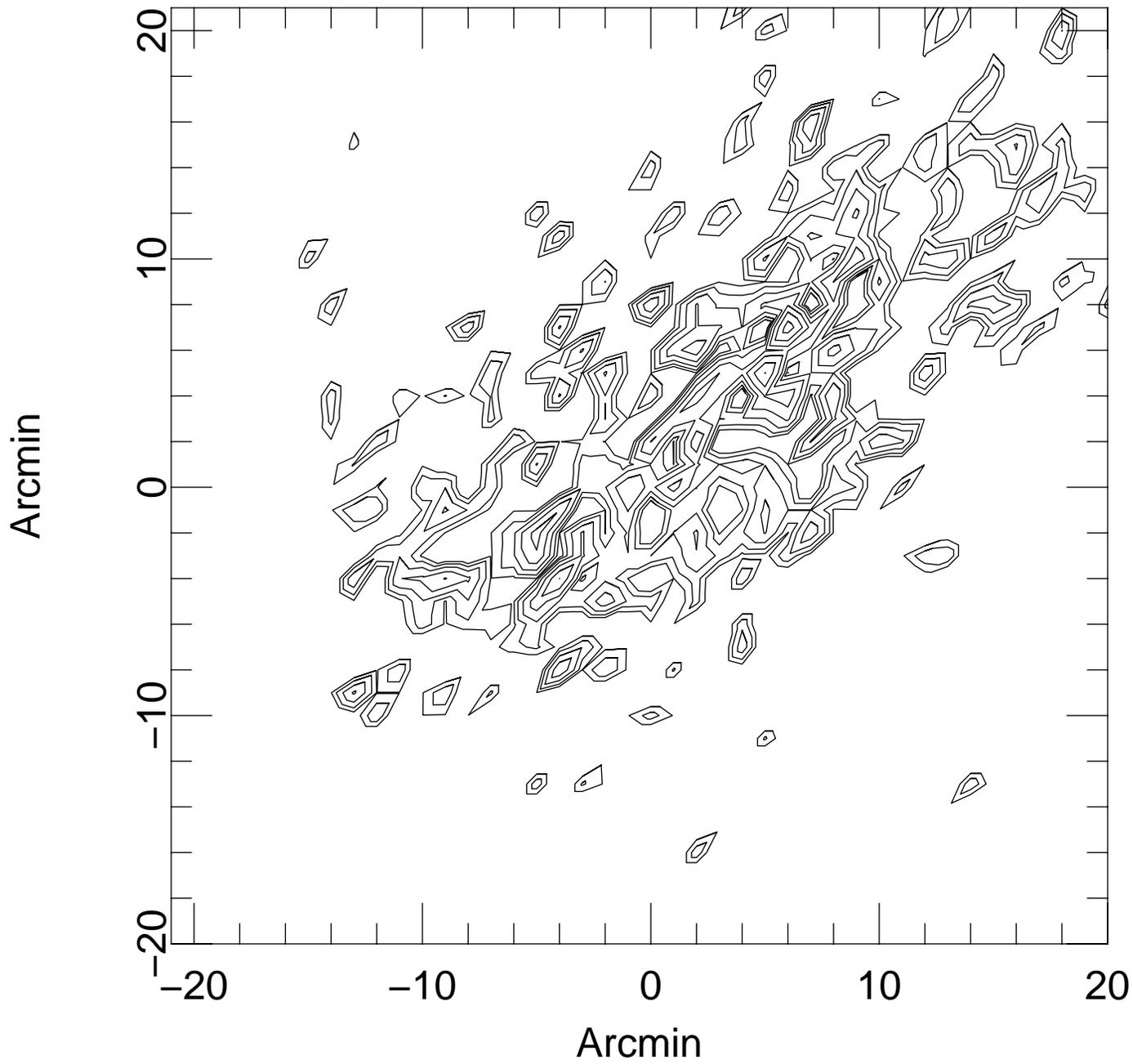
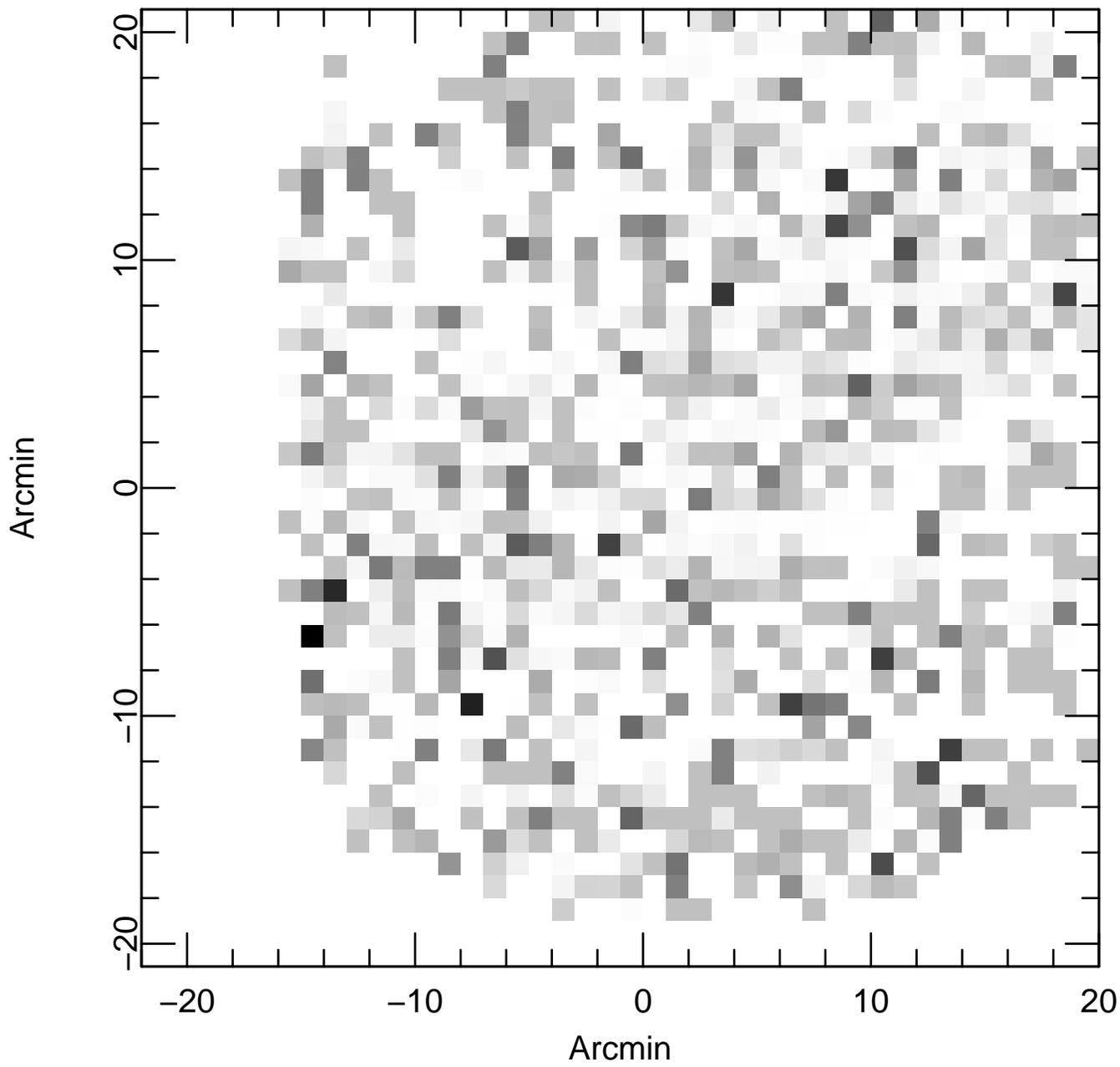
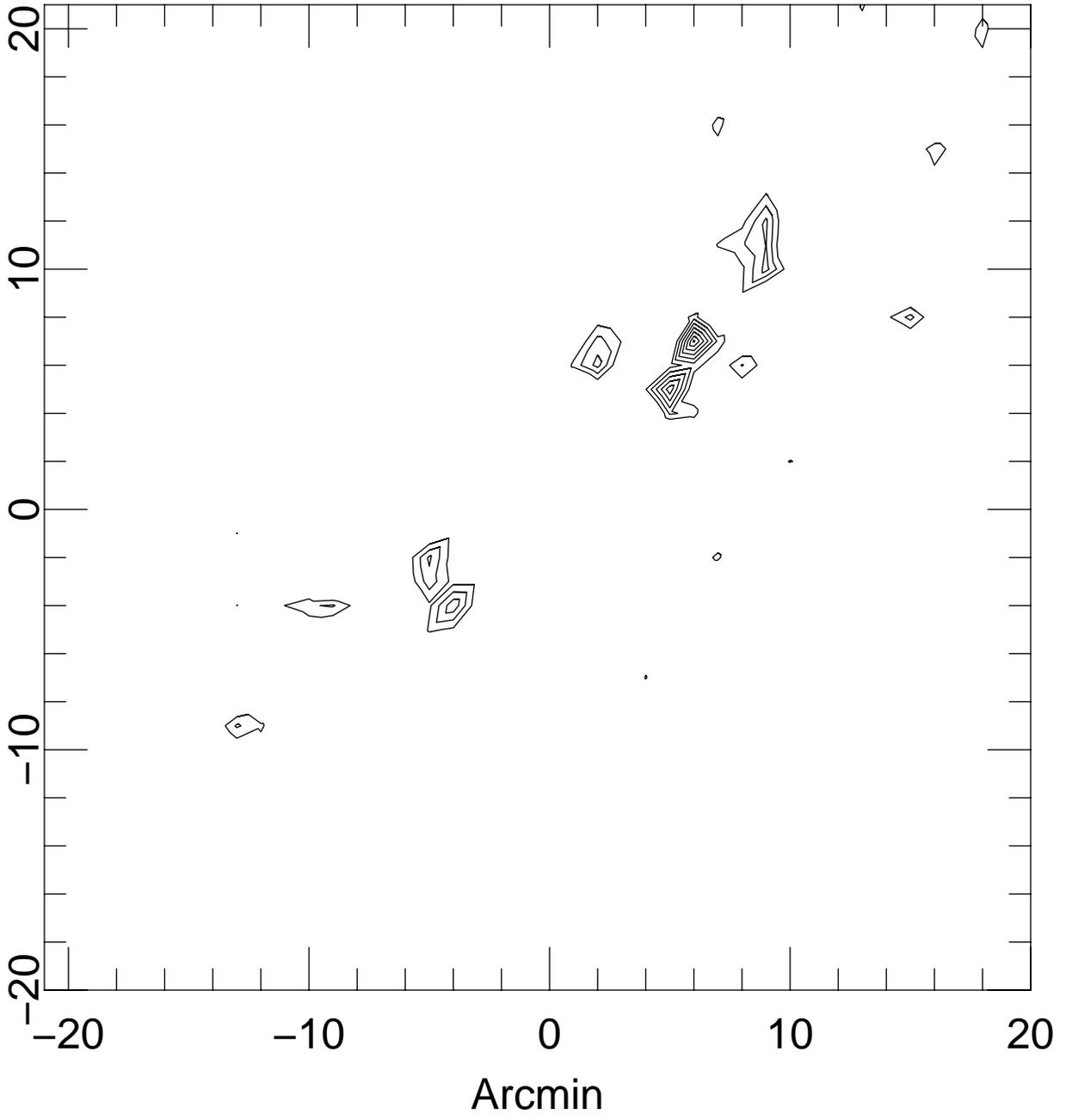
\end{document}